\documentclass[aps,prb,reprint,superscriptaddress,twocolumn,showpacs]{revtex4-1}

\usepackage[T1]{fontenc}
\usepackage[utf8]{inputenc}
\usepackage{textcomp}
\usepackage{color}
\usepackage{comment}
\usepackage{braket}
\usepackage{amsmath}
\usepackage{array,mathtools,amssymb,booktabs}
\usepackage{multirow}

\usepackage{gensymb}

\usepackage{xcolor}
\definecolor{green2}{RGB}{50,160,50}

\usepackage{hyperref}
\hypersetup{
  colorlinks	=true,
  linkcolor	=black,
  citecolor	=blue,
  urlcolor	=blue
}

\newcommand{\muB}{\mu_\mathrm{B}}
\newcommand{\vF}{v_\mathrm{F}}
\newcommand{\EF}{E_\mathrm{F}}
\newcommand{\ED}{E_\mathrm{D}}
\newcommand{\EexRKKY}{E_\mathrm{ex}^\mathrm{RKKY}}
\newcommand{\Eexconv}{E_\mathrm{ex}^\mathrm{conv}}
\newcommand{\EexconvAA}{E_{\mathrm{ex},AA}^\mathrm{conv}}
\newcommand{\EexconvAB}{E_{\mathrm{ex},AB}^\mathrm{conv}}
\newcommand{\Eexconvbri}{E_{\mathrm{ex},\text{bri}}^\mathrm{conv}}

\newcommand{\vR}{\mathbf{R}}
\newcommand{\Eexres}{E_\mathrm{ex}^\mathrm{res}}

\newcommand{\Vg}{V_\mathrm{g}}

\newcommand{\oM}{\overline{M}}
\newcommand{\oDQ}{\overline{\Delta Q}}

\newcommand{\AAmath}{\text{\AA}}

\newcommand{\psip}     { \psi_{ p, \perp             }}
\newcommand{\psipup}   { \psi_{ p, \perp; \uparrow   }}
\newcommand{\psipdown} { \psi_{ p, \perp; \downarrow }}
\newcommand{\psipp}    { \psi_{ p, \parallel                 }}

\newcommand{\psippdown}{ \psi_{ p, \parallel; \downarrow     }}

\newcommand{\Cadyksi} {{C$^\text{ad}_1$}}
\newcommand{\Cadkaksi}{{C$^\text{ad}_2$}}

\DeclareMathOperator{\sign}{sign}

\begin{document}

\title{Gate-tunable magnetism of C adatoms on graphene}

\author{J. Nokelainen}
\email[]{johannes.nokelainen@lut.fi}
\affiliation{LUT University, P.O. Box 20, FI-53851, Lappeenranta, Finland}
\author{I. V. Rozhansky}
\affiliation{Ioffe Institute, 194021 St. Petersburg, Russia}
\affiliation{LUT University, P.O. Box 20, FI-53851, Lappeenranta, Finland}
\author{B. Barbiellini}
\affiliation{LUT University, P.O. Box 20, FI-53851, Lappeenranta, Finland}
\affiliation{Physics Department, Northeastern University, Boston, Massachusetts 02115, USA}
\author{E. L\"ahderanta}
\affiliation{LUT University, P.O. Box 20, FI-53851, Lappeenranta, Finland}
\author{K. Pussi}
\affiliation{LUT University, P.O. Box 20, FI-53851, Lappeenranta, Finland}

\date{\today}

\begin{abstract}
We have performed density functional theory calculations of graphene decorated with carbon adatoms,
which bind at the bridge site of a C--C bond.
Earlier studies have shown that the C adatoms have magnetic moments
and have suggested the possibility of ferromagnetism with high Curie temperature.
Here we propose to use a gate voltage to fine tune the magnetic moments from zero to 1\,$\muB$
while changing the magnetic coupling from antiferromagnetism to ferromagnetism and again to antiferromagnetism.
These results are rationalized within the Stoner and RKKY models.
When the SCAN meta-GGA correction is used,
the magnetic moments for zero gate voltage are reduced and the Stoner band ferromagnetism is slightly weakened in the ferromagnetic region.
\end{abstract}


\maketitle

\section{Introduction \label{sec:intro}}

A functionalization of semiconductor devices with ferromagnetic properties is one of the greatest challenges of
modern spintronics.\cite{Wolf2001_spintronics_review,Zutic2004_spintronics_review,Feng2017_review_2Dspintronics}
Much research is nowadays focused on incorporating magnetic properties into semiconductor system and,
in particular,
into graphene.\cite{Terrones2010_graphene_review,Tucek_2018_review_spintronics,Wang2009,Zhou2009,Giesbers2013,Liu2013a,Sarkar2014,Chen2015a,Makarova2015,
GonzalezHerrero2016,Li2016,Miao2016,Blonski2017,Miao2017,Tucek2017}
Graphene modified by defects is very promising for such purpose.
Because of the reduced coordination number in the two-dimensional system,
the defect states are naturally expected to have weak coupling to graphene,
thus broadening of their energy levels is expected to be small and with high density of states (DOS).
If the Fermi level ($\EF$) further falls inside the broadened energy level,
the Stoner instability\cite{Pavarini2012_kirja} removes the spin degeneracy of the impurity states,
leading to the onset of magnetization.
Thus,
graphene can magnetize even first row element adatoms\cite{Pasti2018}
producing two-dimensional $d^0$ semiconductor magnetism.
This kind of scenario becomes particularly interesting for external gate voltage ($\Vg$) control of the Fermi energy.
In fact,
the impurity state can be easily occupied/drained using $\Vg$.\cite{Chan2011_gated_adatoms,Chan2011_gate_alchemy,Brar2011}
Consequently,
a rich phenomenology emerges from the interplay between magnetic impurities and the gate-voltage control of the interactions.
As we shall see in the next sections,
the description of this interplay is rather complex and it has not been carefully examined in earlier studies.
Experimentally gate-voltage control has been used for producing magnetism on graphene oxide,\cite{Chen2015a}
N-doped graphene oxide\cite{Blonski2017} and graphene grafted with Pt-porphyrins,\cite{Li2016}
revealing the existence of ferromagnetic phases with significant magnetic moments for particular values of $\Vg$.
The magnetic impurities develop a narrow band,\cite{Alexander1964_localized_state_interaction,Pavarini2012_kirja}
which has been proposed to give rise to Stoner band ferromagnetism with high Curie temperatures
even for $sp$ electron systems.\cite{Edwards2006_sp_stoner_ferromag}
Moreover,
the indirect exchange interaction of impurities on graphene has been intensively studied within the
Ruderman-Kittel-Kasuya-Yosida (RKKY)
framework\cite{Power2013_RKKY_review,Saremi2007_RKKY,Brey2007_RKKY,BlackSchaffer2010_RKKY,Sherafati2011a,Sherafati2011b,Kogan2011}
and the recently developed extension of this approach,
which takes into account a resonant hybridization between the adatoms and graphene.\cite{Krainov2015}

Most of the literature considers impurities interacting with one of the two sublattices $A$ or $B$ of graphene.
The Lieb's theorem\cite{Lieb} for bipartite lattice applies for these cases.
As a corollary of this theorem,
two impurities connected to the same sublattice interact ferromagnetically,
while the interaction is antiferromagnetic for impurities connected to different sublattices.
This sublattice-dependence has been verified by the various computational and
experimental studies.\cite{Casolo2009_H-adatoms,Crook2015,GonzalezHerrero2016,Thakur2017,Tucek2017}
For adatoms,
the situation depicted above corresponds to the top-site binding.
However,
the bridge site is in general energetically more favorable than the top site,\cite{Nakada2011,Pasti2018}
despite the fact that it has not been studied as intensively.
  For example, 
  for carbon adatom the adsorption energies to top and hollow high-symmetry sites
  are $0.72$\,eV and $1.36$\,eV higher.\cite{Zhou2011}
Interestingly,
the magnetic interactions for bridge site adatoms are different since coupling with both sublattices occurs,
as pointed out by Gerber \emph{et al.}\ for carbon adatoms.\cite{Gerber2010}
In this situation,
the Lieb's theorem does not apply anymore because the lattice is no more bipartite.
These conditions are therefore more favorable for ferromagnetism within the C adatom network.

\begin{figure}
\includegraphics[width=\linewidth]{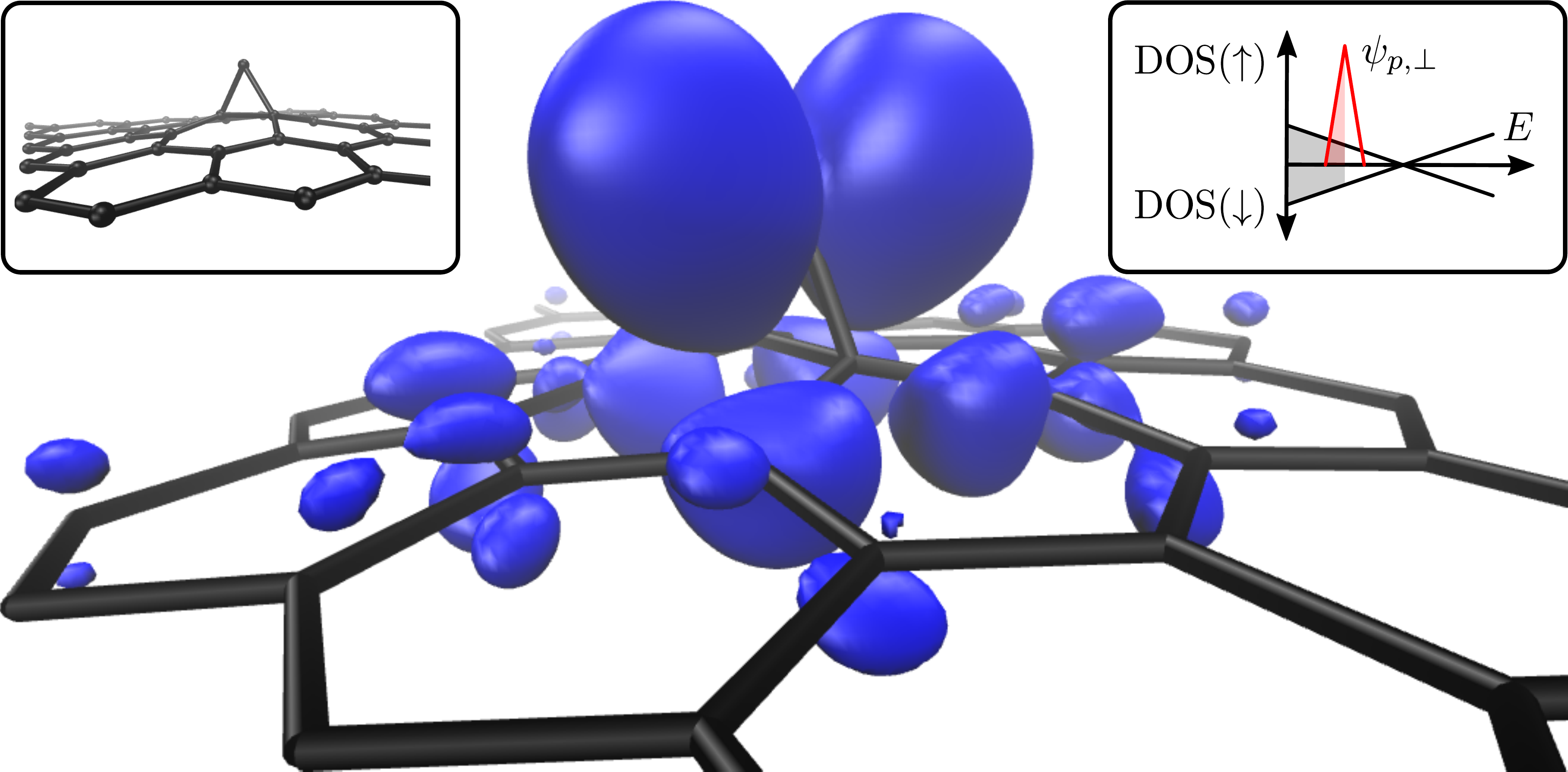}
\caption{\label{fig:vmd}
  (Color online).
  Magnetization density
  isosurface plot for a C adatom on graphene.
  The isosurface is at $0.004\,e/\AAmath^3$,
  while the corresponding negative isosurface is not visible.
  The $\psip$ state (with $p_x/p_y$-symmetry) is visible at the adatom.
  Its $p_z$ character on the graphene sheet can be seen as well.
  Left inset: another view of the geometry.
  Right inset:
  schematic electronic structure,
  the occupied portion of the graphene/$\psip$ states are in gray/red, respectively.
}
\end{figure}

The carbon adatom\cite{Lehtinen2003,Krasheninnikov2004,Singh2009,Fan2010,Gerber2010,
Zhou2011,Ataca2011,Ataca2011a,Ozcelik2013,Kim2014,Hou2015_N_doping_C_migration,Banhart_ACS_2011_graphene_defect_review}
can be considered as an exemplar case for the bridge-site binding.
It has only one partially filled magnetic orbital at about 0.3\,eV below graphene Dirac point ($\ED$).
Here,
we name this orbital as $\psip$ since it has $p$ symmetry with its symmetry axis perpendicular to the C--C bond of the bridge site,
as shown in Fig.~\ref{fig:vmd}.
The $\psip$ orbital hybridizes only weakly with the graphene backbone,
thus it has high partial DOS and it preserves the Dirac cone shape.
Therefore, the ideal model considering a localized state and a Dirac cone remains valid.
Taking all these considerations into account,
one can conclude that C adatoms are interesting for both conceptual and applied purposes.
Nevertheless,
we must bear in mind that the controlled production of C adatom networks on graphene remains a major challenge.
To address these difficulties,
a recent paper by Kim \emph{et al.}\ has suggested a facile pathway for the realization of this system.\cite{Kim2014}
In this study the C adatoms have been monitored using a powerful Raman technique.\cite{Eckmann2012_raman_graphene_sp3}
  As documented by
  a recent review by Banhart \emph{et al.},\cite{Banhart_ACS_2011_graphene_defect_review}
  another obstacle resides in the stability of the adatoms,
  which are mobile even at room temperature due to the rather low migration
  barrier
  of 0.35\,eV--0.60\,eV.\cite{Lehtinen2003,Gerber2010,Ataca2011a,Nakada2011,Ozcelik2013,Hou2015_N_doping_C_migration,Obodo2016_Udo_Schwingenschlogl_C_adatom_barrier}
  Nevertheless,
  intrinsic weak ferromagnetism or superparamagnetism observed in
  graphite\cite{Kopelevich2000,Esquinazi2002,Coey2002,Esquinazi2003,Han2003,Cervenka2009,Saito2011}
  implies that the defects producing this behavior couple to the both sublattices $A$ and $B$ as
  a corollary to the Lieb's theorem.
  Such coupling yields magnetic properties similar to those produced by C adatoms.

The present paper provides a comprehensive theoretical study of graphene decorated with C adatoms
and shows that the magnetic properties can be controlled with a gate voltage.
The paper is organized as follows:
Sec.~\ref{sec:methods} contains the methodology and
part A gives the computational details,
part B contains the used computational supercells
and part C discusses the theoretical framework.
Sec.~\ref{sec:results} reports the results of the study.
Part A focuses on effects of bias to a single adatom
while part B presents the most important results of this study,
which is the behavior of the magnetic interactions as a function of gate voltage for
remote interaction distances and its interpretation.
Part C contains results for close interaction distances.
Part D illustrates the spatial spin polarization patterns induced in graphene by the $\psip$ state.
Sec.~\ref{sec:conclusions} contains the conclusions and outlook of the present work.

\section{Methodology}\label{sec:methods}

\subsection{Computational details}\label{sec:methods:dft}

The calculations were performed within the density functional theory (DFT)
as implemented in the Vienna ab initio simulation package
(VASP).\cite{Kresse1996_VASP_CMS,Kresse1996_VASP_PRB}
  DFT is in principle an exact many-body theory.\cite{Kohn1998_nobel_lecture_DFT}
  However,
  in practice the exchange-correlation (XC) functional taking into account the
  Pauli principle and Coulomb correlation effects contains approximations.
To describe the XC functional,
we used the generalized gradient approximation (GGA)
in the form proposed by Perdew, Burke and Ernzerhof (PBE).\cite{Perdew1996_PBE}
GGA is a correction to the old local density approximation,
which in iron improves the stability of FM phase.\cite{Barbiellini1990}
Moreover,
a set of calculations were repeated with the strongly constrained and appropriately normed (SCAN)
meta-GGA functional,\cite{Sun2015_SCAN}
which is a more precise XC scheme obeying the 17 known exact constraints.
SCAN typically produces superior results compared to most GGA
functionals\cite{Sun2016_SCAN_benchmarking,Car2016_SCAN_Jacobs_ladder,Buda2017_SCAN_thin_film,Furness2018_SCAN_cuprate,Lane2018_La2CuO4_SCAN,Zhang2018_cuprate_stripes}
    including the case of pure graphene,\footnote{
    Buda \emph{et al.},\cite{Buda2017_SCAN_thin_film} discuss the consistency between SCAN and quantum monte carlo 
    results in the pure graphene
} 
  and also our tests on different graphene adatoms\footnote{
    We have tested that,
    unlike PBE,
    SCAN is capable of finding the magnetic moment of F adatoms on graphene.
    Typically hybrid functionals are needed for this result\cite{Kim2013_hybrid_2013}
  }
  support this trend.
  Moreover, 
  Black-Schaffer\cite{BlackSchaffer2010_RKKY} has demonstrated that electron correlation effects 
  play an important role in the coupling of magnetic impurity moments in graphene. 
  Therefore, 
  it is important to check the impact of correlation effect beyond the GGA. 
  These important arguments have justified the deployment of the SCAN functional in our study. 

  The magnetic interactions between C adatoms were studied by placing two adatoms
  on a graphene supercell (SC) and by comparing the energies of the parallel and antiparallel
  adatom spin configurations,
  which we also refer to ferromagnetic (FM) and antiferromagnetic (AFM) solutions throughout the present paper.
  The energy difference
  $\Delta E = E(\text{FM}) - E(\text{AFM})$
  contains the information about magnetic interactions.
  The same SC methodology (using periodic boundary conditions)
  has been successful in describing
  experiments with interacting hydrogen adatoms on graphene.\cite{GonzalezHerrero2016}

The Kohn-Sham orbitals were expanded in a plane wave basis set with an energy cutoff of 600\,eV.
The electron-ion interactions were taken into account using
the projector augmented wave (PAW) method\cite{Blochl1994_PAW,Kresse1999_PAW}
and the electronic energy minimization was performed with a tolerance of 10$^{-5}$\,eV.
Each structure was optimized until the residual forces became
smaller than $0.01\,\text{eV}/\text{\r{A}}$.
In these relaxation runs we used first order Methfessel-Paxton smearing with width of $0.1$\,eV.
The Brillouin zone was sampled with dense $\Gamma$-centered meshes
with $\mathbf{k}$-point separations lower than $0.017\cdot2\pi/\AAmath$.
For example,
a $4\times4\times1$ mesh was used for calculations on $7\times7$ repeated graphene SCs and
a $4\times8\times1$ mesh was used for $8\times4$ SCs.
We found that the high $k$-point density is essential in describing the magnetic $\psip$ state of
the C adatom correctly.
Some of these parameters were refined for the SCAN simulations
(see Sec.~1.1 of the supplemental material\cite{Supplemental}).
To obtain more accurate total energies and DOS,
the Methfessel-Paxton method was upgraded to the tetrahedron method with
Bl\"ochl corrections.\cite{Blochl1994_ismear-5}
Moreover,
the $\mathbf{k}$-point grids were increased to
$10\times10\times1$ mesh for $N\times N$ SCs and
$8\times16\times1$ mesh for $2N\times N$ SCs.
As in earlier work,\cite{Chan2011_gated_adatoms,Chan2011_gate_alchemy,Suarez2011,Sofo2011}
the gate voltage was modeled by adding $\Delta Q/e$ electrons.
In order to avoid divergence in the Ewald summation,
the unit cell was kept charge neutral
by adding a compensating jellium background charge.
These computational schemes may produce errors as discussed in detail
in Sec.~1.2 of the supplemental material.\cite{Supplemental}
Moreover,
when $\Delta Q\geq1.5\,e$ and when the SCs are small,
some charge spilling to the vacuum occurs.
This problem was overcome by decreasing the the unit cell height
to $15\,\AAmath$ or even $13\,\AAmath$ from the standard value of $20\,\AAmath$.
The forementioned errors are increased by the decreasing of the unit cell height,
but our results are not significantly affected by all these errors as
explained in the Sec.~1.2 of the supplemental material.\cite{Supplemental}

\subsection{Studied configurations}\label{sec:methods:confs}

We studied the magnetic interactions of the two adatoms \Cadyksi\ and \Cadkaksi\ on
a set of different geometrical configurations.
These configurations are characterized by
adatom orientations,
graphene SCs (on which the adatoms are placed)
and one-sided/two-sided adsorption sites,
where the adatoms are adsorbed on the same/opposite sides of the graphene sheet.
Two main configuration types were used,
as explained by the following paragraphs.

\begin{figure}
\includegraphics[width=\linewidth]{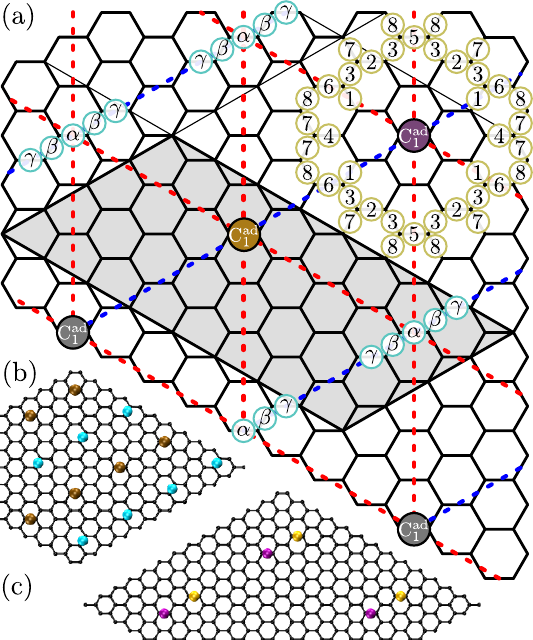}
\caption{
  (Color online).
  (a): In the adatom pairs the purple {\Cadyksi} is the first adatom
  and the second one is located in one of the sites labeled \#1 to \#8.
  In the adatom arrays the ochre {\Cadyksi} represents the first adatom in the unit cell 
  (the gray area) and the other {\Cadyksi} are its periodic images.
  The second adatom and its periodic images are located at sites $\alpha$, $\beta$ or $\gamma$.
  The present illustration is for the SC size $8\times4$.
  The dashed lines indicate nearest-neighbour interactions for the $\alpha$ array.
  All the symmetry-equivalent positions of {\Cadkaksi} are shown.
  (b): The $\beta(6\times3)$ array, where the unit cell is repeated $2\times4$ times.
  (c): Configuration~\#6 on a $7\times7$ SC, where the unit cell is repeated $2\times2$ times.
}
\label{fig:unitcell}
\end{figure}

In \emph{adatom pair configurations},
the adatoms were placed close to each other.
These configurations were used to model short range pairwise interactions
as in the study by Gerber \emph{et al.}\cite{Gerber2010}
The results for these pairs are discussed in Sec.~\ref{sec:results:close}.
Eight different orientations of the adatoms were considered,
labeled \#1 to \#8,
and they are illustrated in Fig.~\ref{fig:unitcell}.
These configurations contain all the pairs that are separated at most $5.8\,\AAmath$ (Conf.~\#8)
and are stable for one-sided adsorption.
In Conf.~\#1 the adatoms are separated by 3.8\,\AA,
and for shorter separations the adatoms dimerize.\cite{Ataca2011a,Gerber2010}
The dimerized solution is further discussed in Sec.~2 of the supplemental material.\cite{Supplemental}
The pair configurations were studied in two supercell sizes.
The first is the $7\times7$ SC
containing 98 graphene atoms plus the two adatoms.
In this case,
the separations between periodic images
are at least $11.7\,\AAmath$ (Conf.~\#8).
As discussed in the Sec.~3.1 of the supplemental material,\cite{Supplemental} 
these separations guarantee reasonably good isolation of the adatom pairs.
The SC size $7\times7$ is also used by Gerber \emph{et al.},\cite{Gerber2010}
and their Confs.~C, D and E correspond to our Confs.~\#3, \#2 and \#6,
respectively.
Another SC size is $6\times3$,
these SCs have 36 graphene atoms plus the two adatoms.
The interactions between adatoms and their periodic images are strong.
Thus the adatom pairs interact with their surroundings.
These cells were intended for simulating random adatom distributions
with a coverage of one adatom per 18 graphene atoms.

In \emph{adatom array configurations},
the adatoms and their periodic images form regular and infinite arrays,
which were employed in simulations involving long-range interactions.
These arrays replace the isolated adatom pairs,
which would have prohibitively large SCs in the case of remote interactions.
The adatom array results are discussed in Sec.~\ref{sec:results:remote}.
The studied array types were labeled $\alpha$, $\beta$ and $\gamma$.
For $\alpha$ arrays a $N \times N$ graphene SC with one adatom
was repeated twice to yield two adatoms in a $2N \times N$ SC.
This operation leads to equilateral triangular array
as illustrated in Fig.~\ref{fig:unitcell}\,(a) for the case of $8\times4$ SC.
In this study,
two SC sizes ($6\times3$ and $8\times4$)
were used and they correspond to {$\sim$\,7.5\,\AA} and {$\sim$\,10\,\AA} adatom separations,
respectively.
For the $\beta$ arrays,
{\Cadkaksi} was moved to the neighboring bridge site,
breaking the $D_{3h}$ symmetry of the $\alpha$ arrays.
The $\gamma$ arrays were obtained by further moving {\Cadkaksi}.
In these
arrays six nearest-neighbour adatom-adatom interactions per unit cell are present.
However,
in interactions between an adatom and its own periodic images the spins are parallel in both FM and AFM solutions,
hence their contribution to $\Delta E$ cancels.
There are two interactions per unit cell leading to this cancellation
(blue dashed lines in Fig.~\ref{fig:unitcell}\,(a)),
hence only four nearest-neighbor interactions
(red dashed lines in Fig.~\ref{fig:unitcell}\,(a))
contribute to $\Delta E$.

\subsection{Theoretical analysis}\label{sec:methods:hamiltonian}

In conventional RKKY theory,
the indirect exchange interaction between two adatoms mediated by delocalized electrons is described by the following Hamiltonian:\cite{Kittel-book}
\begin{equation}
\label{eqRKKYHam}
\hat{H}_{\mathrm{RKKY}} = J \delta({\bf r} - {\bf R}_1) \hat{\bf S} {\bf I}_1 + J \delta({\bf r} - {\bf R}_2) \hat{\bf S} {\bf I}_2,
\end{equation}
where $\hat{\bf S}$ is the spin operator of an electron mediating the exchange interaction,
${\bf I}_{1,2}$ are the adatoms spins, $J$ is the exchange constant, ${\bf R}_{1,2}$
denote positions of the two adatoms.
The energy of the system is obtained by treating Eq.~\eqref{eqRKKYHam} as a perturbation,
the second order correction depends on the configuration of the adatoms spins.
This interaction oscillates as a function of the distance between the magnetic centers,
in the case of graphene its sign also depends on the positions of the adatoms
in relation to the graphene sublattices $A$ and $B$.
For small adatoms separations the alignment is FM for $AA$ and AFM for $AB$ configuration.
This well-known result is a consequence of Lieb's theorem for bipartite lattice.\cite{Lieb}
A completely different situation occurs if the bound adatom states at energy $\varepsilon_0$
are resonantly coupled to the graphene 2D continuum,
i.e.,
$\varepsilon_0$ falls in the range of occupied states of graphene ($\varepsilon_0\ll\EF$).
\cite{Krainov2015}
In this case an effective resonant hybridization occurs between the bound states and the graphene states
lying within a small energy range near $\varepsilon_0$.
The resonant type of hybridization makes the perturbation approach used by conventional RKKY theory inapplicable.
However,
the problem can be solved using a different approach.\cite{Rozhansky2015_JMMM,Krainov2015,Rozhansky2014_RKKY_PSS}
This generalized theory of indirect exchange considers the following Hamiltonian:
\begin{equation}\label{Htot}
 \hat{H} = \hat{H}_0  + \hat{H}_\mathrm{T}  + \hat{H}_\mathrm{RKKY},
\end{equation}
where $\hat{H}_0$ describes the non-interacting adatoms and graphene,
$H_\mathrm{T}$ describes the coupling of the magnetic centers to the graphene,
in particular it incorporates the details of the coupling to the sublattices $A$ and $B$,
$\hat{H}_\mathrm{RKKY}$ is the exchange term of Eq.~\eqref{eqRKKYHam}.
The general expression (valid for arbitrary $\EF$)
for the indirect exchange energy reads:\cite{Krainov2015}
\begin{equation}\label{eqExGen}
  \EexRKKY
  =
  \int\limits_{ - \infty }^{{\EF}}
  {\frac{{\mathrm{d}E}}{\pi }}
  \arctan \frac{{{j ^2}{E^2}g(E,\vR)\sign{E}}}{{\big[ {{{({\varepsilon _0} - E)}^2} - {j ^2}} \big]^2}}.
\end{equation}
Here
$j$ is the exchange energy constant describing direct exchange interaction between the adatom and graphene.
The function $g(E,{\vR})$ encapsulates all the details of the coupling between two interacting adatoms and graphene
and ${\vR}$ connects the adatoms.
In the case of the resonant coupling ($\varepsilon_0\ll\EF$),
only the energy range close to $\varepsilon_0$ contributes to the interaction
due to the poles of the integrand in Eq.~(\ref{eqExGen}).
In other words,
the resonant indirect exchange ($\Eexres$) is effectively mediated only by the electrons with approximately resonant energy $\varepsilon_0$
and the adsorption geometry does not play any role.
Consequently,
the sign of $\Eexres$ only depends on the position of the resonant state relative to the Dirac point:
\begin{equation}\label{eq:ResRKKY}
  \sign \Eexres= - \sign \left( \varepsilon_0 - \ED \right).
\end{equation}
If the bound impurity state lies above $\ED$,
the indirect exchange is mediated by electron-like states of graphene
and the indirect exchange is FM at a small distance between the adatoms,
whereas if $\epsilon_0<\ED$,
it is AFM at a small distance
as it is mediated by graphene hole-like states.\cite{Krainov2015}
In the case of carbon adatoms,
the $\psip$ state is located below $\ED$,
therefore,
we expect the C adatoms to interact antiferromagnetically
when $\EF$ is tuned above $\epsilon_0$ by gate voltage,
which we refer to as \emph{resonant region}.

\begin{table}
\caption{
  Dependency of the sign of $\EexRKKY$ to geometry of the system and
  to location of $\epsilon_0$ with respect to $\EF$ and $\ED$.
}
\begin{ruledtabular}
\begin{tabular}{cccccc}
&
& \multicolumn{2}{c}{\multirow{2}{*}{Conventional}}
& \multicolumn{2}{c}{\multirow{2}{*}{Resonant}}
\vspace{6pt}
\\
&
& \multicolumn{2}{c}{RKKY ($\EF<\epsilon_0$)}
& \multicolumn{2}{c}{RKKY ($\EF>\epsilon_0$)}
\\
\cline{3-4}
\cline{5-6}
\vspace{-8pt}
\\
&
&$\epsilon_0<\ED$
&$\epsilon_0>\ED$
&$\epsilon_0<\ED$
&$\epsilon_0>\ED$
\\
\multicolumn{2}{c}{Geometry}
&\includegraphics[scale=.7]{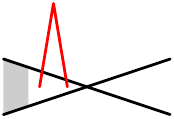}
&\includegraphics[scale=.7]{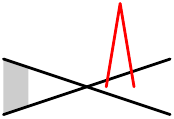}
&\includegraphics[scale=.7]{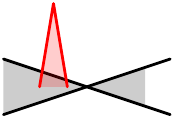}
&\includegraphics[scale=.7]{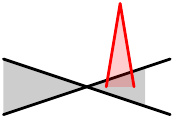}\\
\colrule
$AA$ & \includegraphics[scale=.7]{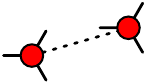} & FM  &  FM & AFM & FM \\
$AB$ & \includegraphics[scale=.7]{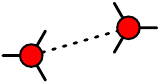} & AFM & AFM & AFM & FM \\
Bri  & \includegraphics[scale=.7]{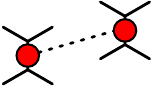} & AFM/FM & AFM/FM & AFM & FM \\
\end{tabular}
\end{ruledtabular}
\label{tab:RKKY}
\end{table}

The opposite case ($\varepsilon_0\gg\EF$) is the familiar conventional RKKY.
Mathematically the difference stems from the poles being outside the integration limits in Eq.~\eqref{eqExGen},
thus the whole range of the occupied states contributes to the integral.
For undoped graphene, 
Eq.~(\ref{eqExGen}) simplifies for the $AA$, $AB$ and bridge-bridge configurations into
\begin{align}\label{eq:Enonres}
\EexconvAA(\vR)
&=
  -\frac{{\tau {j^2}}}{{8\pi \varepsilon _0^4}}
  {\left( {\frac{{\hbar {\vF}}}{R}} \right)^3}
  {\cos ^2}\left( {\frac{{x(\vR)}}{2}} \right),
\nonumber
\\
\EexconvAB(\vR)
&=
  \frac{{3\tau {j^2}}}{{8\pi \varepsilon _0^4}}
  {\left( {\frac{{\hbar {\vF}}}{R}} \right)^3}
  {\cos ^2}\left( {\frac{{ x(\vR)}}{2} - {\theta_\vR}} \right),
\nonumber \\
\Eexconvbri(\vR)
&\approx
\EexconvAA(\vR_{AA}) + \EexconvAA(\vR_{BB})
\nonumber\\
&+\EexconvAB(\vR_{AB}) + \EexconvAB(\vR_{BA}),
\end{align}
where $R=\|\vR\|$,
$\theta_\vR$ is the polar angle of $\vR$,
$x(\vR)=\left({{\bf{K}} - {{\bf{K}}^\prime }}\right)\cdot\vR$,
${\bf K}$ and ${\bf{K}}^\prime $ are the graphene Dirac points,
$\tau$ is the energy parameter defining the strength of the adatoms coupling to graphene and
$\vR_{AB}$ is the vector connecting the $A$ sublattice basal site of the first adatom
to the $B$ sublattice basal site of the second adatom.\footnote{
  For doped graphene analytical formulae have been considered in Ref.~\onlinecite{Sherafati2011b}
}
Eqs.~\eqref{eq:Enonres} are in agreement with the conventional RKKY theory results for graphene,\cite{Power2013_RKKY_review,
Saremi2007_RKKY,Brey2007_RKKY,BlackSchaffer2010_RKKY,Sherafati2011a,Sherafati2011b,Kogan2011}
the difference in the pre-factor is due to the details of the model
where the bound state level exists also in a non-resonant case.\cite{Krainov2015}
As a consequence of Eqs.~(\ref{eq:Enonres}),
the $AA$ interaction is FM while $AB$ is AFM.
However,
the sign of $\Eexconvbri$ depends on the relative position of the adatoms.
It can be approximated by summing the four $AA$/$AB$ type interaction pairs between
the two basal sites of the first adatom and the two basal sites of the second adatom,
as in Ref.~\onlinecite{Sherafati2011a} for selected configurations.
As the amplitude of the AFM terms is three times larger than the FM contributions,
AFM configurations are more frequent.
For example,
if $\theta_\vR = \pi/2$,
$\Eexconvbri$ is FM only if the separation between adatoms is $3n$ graphene unit cells
($n\in\mathbb{N}$),
otherwise it is AFM.
The $n=2$ case corresponds to Conf.~\#5 defined in Sec.~\ref{sec:methods:confs}.
For the adatom arrays
$\Delta E$ can be approximated by summing the nearest neighbor interactions.
For example,
in the $\alpha(8\times4)$ array shown in Fig.~\ref{fig:unitcell},
there are two different adatom pairs:
(i) $\theta_\vR=\pi/2$, $n=4$, (ii) $\theta_\vR=\pi/6$, $n=4$.
The sum of their contributions is AFM.
The results on both conventional and resonant RKKY are summarized in Table~\ref{tab:RKKY}. 
The results are valid under the rigid band approximation.
Effects beyond this approximations have been discussed by Shiranzaei \emph{et al.}\cite{Shiranzaei2018_PRB_nonlinear_spin_suspectibility_in_topological_insulators}

Both RKKY theories discussed above assume the non-polarized continuum
of the mobile carriers mediating the indirect exchange interaction between the adatoms.
However, in practice the impurity states have some spatial spillover on the graphene backbone
and therefore a hybridization occurs and an impurity band develops with a typical width proportional
to the coupling strength.\cite{Pavarini2012_kirja}
A weak coupling between the impurity states leads to a narrow band with a large DOS.
When the Fermi level being adjusted by $\Vg$ falls inside such an impurity band,
the spin polarization of the mobile carriers becomes favorable leading to the onset
of ferromagnetism according to Stoner model. 
We refer to this region as the \emph{intermediate region}.
In order to study the onset of ferromagnetism,
we consider the
Hubbard Hamiltonian to account for the on-site electron-electron interaction:
\begin{equation}
\label{eq:Hub}
{H_U} = \sum\limits_i {U{n_{i \uparrow }}{n_{i \downarrow }}},
 \end{equation}
where the summation is over all graphene and adatom sites,
$n_{i\sigma}$ is the occupation number operator for electron with spin projection $\sigma$ at site $i$.
Here $U$ is the on-site Coulomb interaction energy for the two electrons occupying carbon $p_z$ atomic orbital.
As the result of the mean-field approximation the Hubbard model leads to the Stoner model,\cite{Pavarini2012_kirja,Tucek_2018_review_spintronics}
which typically yields ferromagnetism when the Stoner criterion is satisfied for the onset of magnetization:
  \begin{equation}
    \label{eq:Stoner}
    U\cdot\text{DOS}(\EF)>1.
  \end{equation}
For graphene without adatoms DOS$(\EF)$ remains at about $1\text{ eV}^{-1}$
for a moderate gate voltage applied. 
However, an impurity band induced by hybridization with the adatoms would have
a substantially larger DOS.
As can be seen in Fig.~\ref{fig:dos}, the DOS in an impurity band induced by the hybridization of the adatoms bound states with graphene exceeds $\sim 10 \text{ eV}^{-1}$.
As for now, there are no direct experiments which allow to extract a value for the model parameter $U$ to adequately
reflect the Coulomb electron-electron interactions in graphene.
In some works,
it is reported to be $U\sim t$,
where $t\approx2.7$\,eV is the tight-binding hopping parameter of graphene.\cite{Yazyev}
Other experiments indicate that the strength of the Coulomb interaction in graphene is
about an order of magnitude smaller.\cite{Reed2010}
Even with
this lower estimate $U\sim0.1$ eV
the Stoner criterion (\ref{eq:Stoner}) is fulfilled
when $\EF$ adjusted by the gate voltage falls in the impurity band.
At that the FM solution becomes energetically favorable in the intermediate region as our DFT calculations confirm.

\section{Results}\label{sec:results}

\subsection{One C adatom per unit cell} \label{sec:singleadatom}

\begin{figure}
\includegraphics[width=\linewidth]{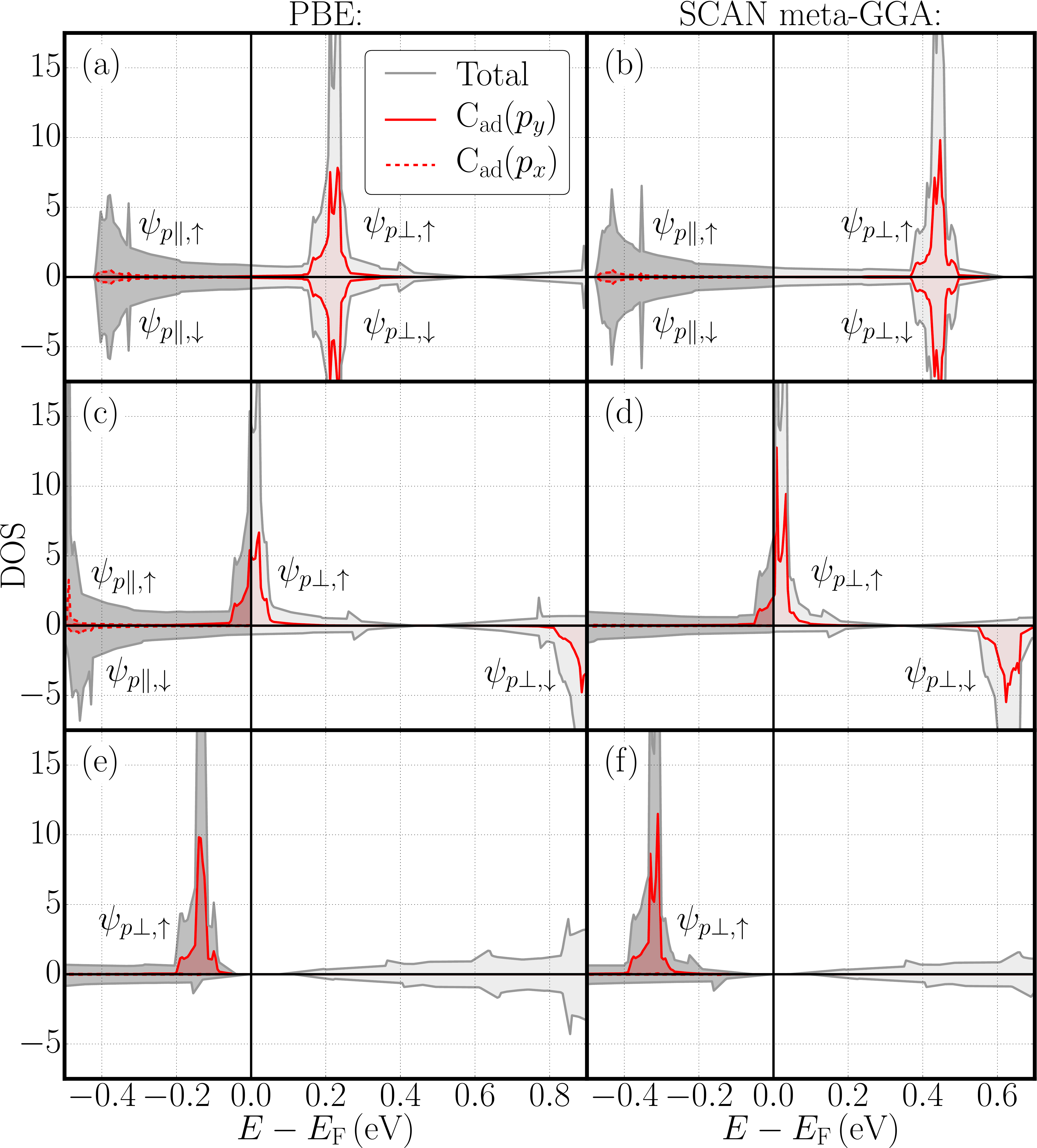}
\caption{
  (Color online).
  Total DOS and $lm$-decomposed PDOS projected to the adatom for $\Delta Q$ values of
  (a): $-0.5\,e$ (PBE).
  (b): $-0.5\,e$ (SCAN).
  (c): $0\,e$ (PBE).
  (d): $0\,e$ (SCAN).
  (e): $1\,e$ (PBE).
  (f): $1\,e$ (SCAN).
  In each case $\EF$ has been set to zero.
  Positive and negative DOS values denote DOS($\uparrow$) and DOS($\downarrow$),
  respectively.
}
\label{fig:dos}
\end{figure}

We first focus on effects of gate voltage to the magnetic and geometrical properties of a single C adatom on $N\times N$ supercells,
which leads to a periodic adatom coverage due to the periodic boundary conditions.
Figure~\ref{fig:dos} illustrates the DOS of such system with a $4\times4$ SC,
as well as $l,m$ quantum number decomposed partial DOS (PDOS) projected to the adatom for different $\Delta Q$ values.
In this PDOS one can see contributions from the magnetic $\psip$ orbitals and also from
strongly hybridized lower-lying adatom states denoted as $\psipp$.
When 0.5 electrons have been removed from the unit cell,
both $\psipup$ and $\psipdown$ orbitals are unoccupied and degenerate in energy
(Figs.~\ref{fig:dos}\,(a) and \ref{fig:dos}\,(b)).
When $\Vg=0$ (Figs.~\ref{fig:dos}\,(c) and \ref{fig:dos}\,(d)),
the $\psipup$ orbital becomes partly occupied and the $\psipdown$ orbital is lifted in energy by Stoner splitting.
Therefore,
the C adatom network magnetizes and the magnetic moment per adatom is
$M=0.38\,\muB$ for PBE and $M=0.24\,\muB$ for SCAN.
The application of a positive $\Vg$ fills the $\psip$ impurity state completely
as illustrated in Figs.~\ref{fig:dos}\,(e) and \ref{fig:dos}\,(f) for the case $\Delta Q = 1\,e$.
Interestingly,
at $\Delta Q = 1\,e$ the Fermi level is exactly at $\ED$ since the extra electron is completely absorbed by the $\psip$ orbital.
Therefore,
only electrons of the graphene matrix are present at the Fermi level.

\begin{figure}
\includegraphics[width=\linewidth]{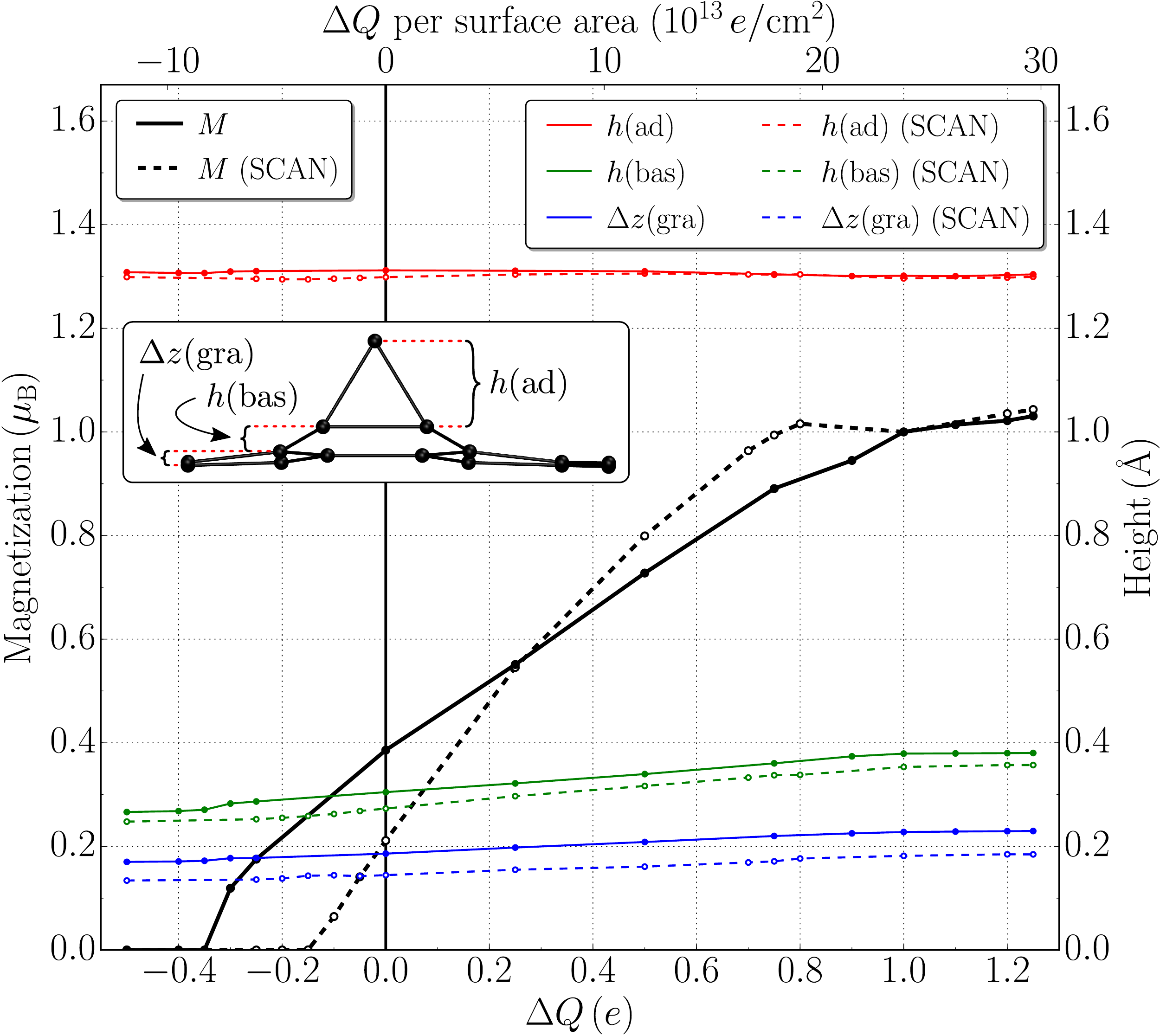}
\caption{
  (Color online).
  Magnetization $M$,
  height of the adatom $h$(ad),
  height of the basal atoms $h$(bas) and
  buckling of the graphene sheet $\Delta z$(gra)
  as a function of $\Delta Q$ for a $4\times4$ SC with one adatom.
  Also $\Delta Q$ per surface area is given in the upper $x$-axis.
  Inset:
  side view of the PBE structure for $\Delta Q = 1.25\,e$ and the definitions for
  the geometrical parameters.
  }
  \label{fig:4x4_mag_geom}
\end{figure}

Comprehensive magnetization results are given by $M(\Delta Q)$ curves presented in Fig.~\ref{fig:4x4_mag_geom}.
For PBE,
the curve is rather linear with a slope of about $0.65\,\muB/e$ until the magnetization saturates at $\Delta Q = 1\,e$.
This slope is rather steep because the PDOS of the $\psip$ state is high compared to the graphene DOS.
However,
in the region below $M\lesssim0.2\,\muB$ the curve presents a nonlinear behavior
related to a breakdown of the Stoner model.
For SCAN significant differences with respect to the PBE results can be noticed.
In particular,
the growth of the magnetization curve is steeper.
Below $M\approx0.5\muB$ the slope is 1.4$\,e/\muB$,
but it gradually decreases to 0.8$\,e/\muB$ before saturation.
Interestingly,
the magnetization slope is larger than $1\,\muB/e$ for most of the curve,
which implies that $\psip$ moves downwards in energy while being filled from the voltage bias,
and therefore also attracts electrons from the graphene lattice.
This feature of SCAN (which contains corrections to PBE)
indicates a stronger response of magnetic moment to $\Vg$ in comparison to PBE.
Another experimentally verifiable SCAN result is that at $\Delta Q = 0\,e$,
the magnetization ($0.24\,\muB$) is significantly reduced in comparison to the PBE result ($0.38\,\muB$).
We have observed similar trend in simulations with other SCs.
This observation
can explain the fact that intrinsic carbon magnetism is observed only in
rare occurrences\cite{Kopelevich2000,Esquinazi2002,Coey2002,Cervenka2009,Saito2011}
instead of being a common phenomenon.
Moreover,
our results indicate that gate voltage can be used as a control tool to enhance magnetism
in various graphitic samples that are either weakly magnetic or even nonmagnetic.
In fact,
the same principle applies to various graphene-based magnetism,
as recently experimentally demonstrated.\cite{Chen2015a,Blonski2017,Li2016}

Our results with other SC sizes show that the slopes of $M(\Delta Q)$ curves become smaller when the adatom is placed on bigger SCs.
This observation can be explained as follows.
For bigger SCs there is a larger number of graphene states.
Therefore,
a larger $\Delta Q$ contribution occupies the graphene states instead of the $\psip$ band.
However,
smaller $M(\Delta Q)$ slopes do not necessarily mean weaker magnetic response to the gate voltage
since this response (i.e., slope of $M(\Vg)$) is related to the DOS peak width of the impurity state.
In fact,
the peak width clearly decreases with increasing SC size because the hybridization
between the localized $\psip$ state with its periodic images is reduced.
The peak widths at $\Delta Q = 0$ have the following trend for different SCs:
$3\times3$: 0.23\,eV;
$4\times4$: 0.10\,eV;
$5\times5$: 0.08\,eV;
$6\times6$: 0.04\,eV;
$7\times7$: 0.03\,eV;
$9\times9$: 0.03\,eV;
$9\times9$: 0.02\,eV.
\footnote{
  The peak widths have been taken from DOS plots with $\EF$ taken above the $\psip$ state by
  positive $\Delta Q$.
}
Nevertheless,
more dilute adatom coverage produces also smaller magnetization per surface area
and weaker magnetic coupling between adatoms.

The gate voltage affects also geometrical properties according to Fig.~\ref{fig:4x4_mag_geom}.
The inset of Fig.~\ref{fig:4x4_mag_geom} defines
$h$(ad) as the height of the adatom C with respect to the basal atoms,
$h$(bas) as the height of the basal atoms with respect to the nearest neighbor graphene atoms and
$\Delta z$(gra) as buckling of graphene.
The total height of the structure is
$\Delta z\text{(tot)}=h\text{(ad)} + h\text{(bas)} + \Delta z\text{(gra)}$.
Without gate voltage,
these values are for SCAN
$ h(\text{ad}) = 1.30\,\AAmath$,
$ h(\text{bas}) = 0.27\,\AAmath$,
$ \Delta z(\text{gra}) = 0.15\,\AAmath$ and
$ \Delta z(\text{tot}) = 1.72\,\AAmath$.
$\Vg$ does not have much effect on $h$(ad) nor in overall on the shape of the nearly equilateral triangle
formed by the adatom and the basal atoms.
More noticeable is the upwards movement of the triangle when $\Delta Q$ is filled.
This motion can be tracked in the increase of $h$(bas) by {0.08\,\AA} and in the increase of $\Delta z$(gra) by {0.04\,\AA}.
The PBE values are nearly identical for $h(\text{ad})$,
about {0.03\,\AA} larger for $h(\text{bas})$ and
{0.04\,\AA} to {0.06\,\AA} larger for $\Delta z(\text{gra})$.
The upwards motion of the triangle and increase in the buckling of graphene,
produced by the filling of $\psip$ state,
could be explained by the repulsive Coulomb interaction of the $\psip$ state
with the $p_z$ orbitals of graphene.
Nevertheless,
$ h(\text{bas})$ and $ \Delta z(\text{gra})$ are significant even when the $\psip$ state is emptied.

We have also verified that our binding energies are consistent with previously calculated values of
$-1.46$\,eV\cite{Pasti2018} and
$-1.63$\,eV.\cite{Singh2009}
Our results yield
$-1.46$\,eV, $-1.52$\,eV, $-1.51$\,eV and $-1.52$\,eV with
$3\times3$, $4\times4$, $5\times5$ and $9\times9$ SCs, respectively.
Therefore,
the binding energies are not sensitive to the SC size until at high concentrations.
For SCAN a slight decrease in binding energy was observed as
for $4\times4$ and $5\times5$ SCs our results gave values of
$-1.44$\,eV and $-1.46$\,eV,
respectively.
However,
our PBE results for the energies of the top and hexagonal special symmetry adsorption sites are lower than in the literature,
which is possibly because we let the graphene fully relax
while constraining the symmetry of these adsorption sites.
For the top site we obtained an energy which is 0.63\,eV higher than for the bridge-site,
meanwhile the reported values are 0.72\,eV\cite{Zhou2011} and 0.86\,eV.\cite{Ataca2011}
For the hexagonal site our value is 1.27\,eV,
and the literature values are 1.36\,eV\cite{Zhou2011} and 1.81\,eV.\cite{Ataca2011}

\subsection{Remote interaction adatom array configurations}\label{sec:results:remote}

\begin{figure*}[t]
  \centering
  \includegraphics[width=0.91\linewidth]{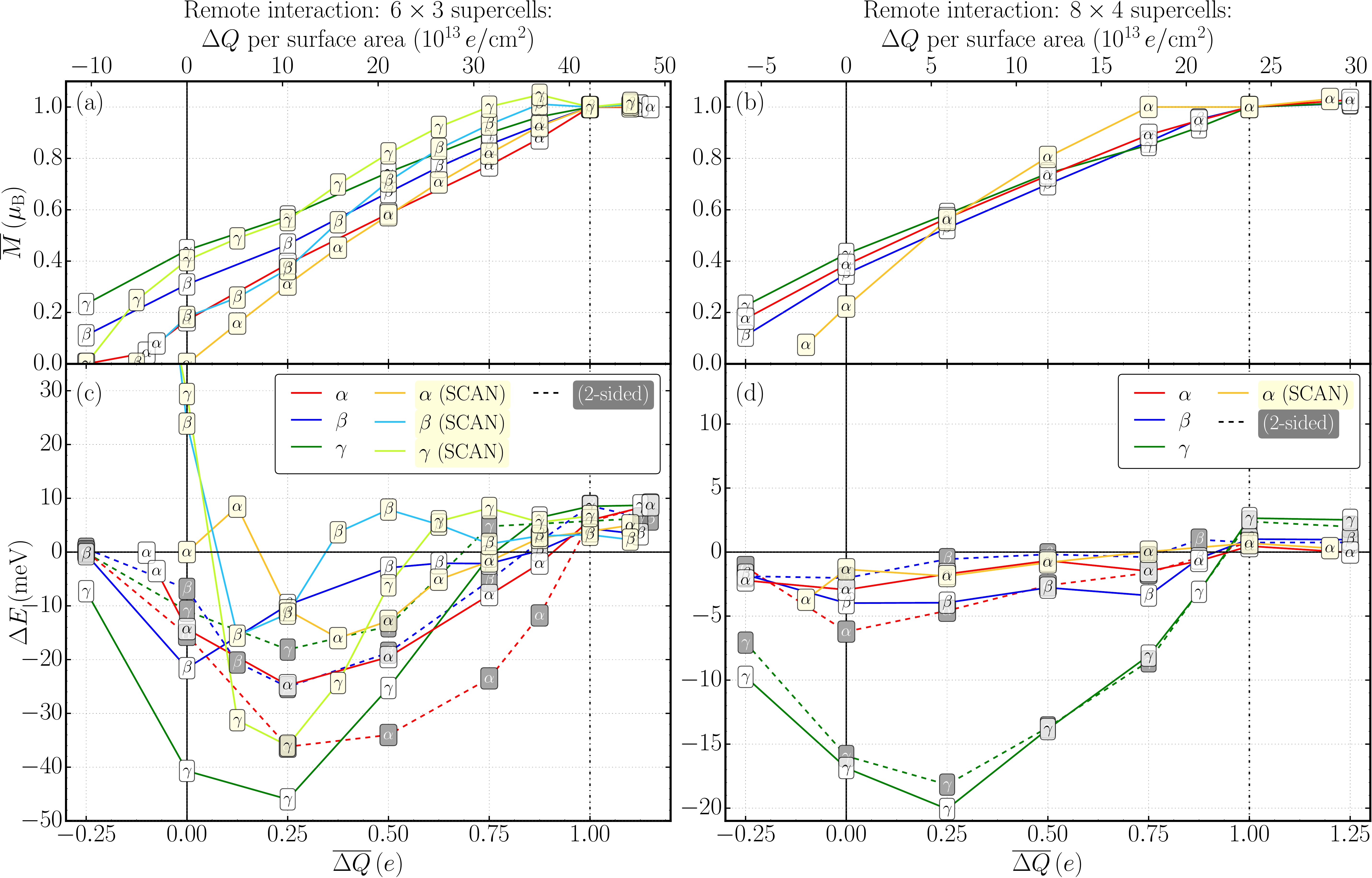}
  \caption{
    (Color online).
    Remote interaction arrays:
    Panels (a) and (b) represent $\oM=M(\text{FM})/2$ for $6\times3$ and $8\times4$ SCs,
    respectively.
    Panels (c) and (d) contain the $\Delta E = E(\text{FM}) - E(\text{AFM})$ values for $6\times3$ and $8\times4$ SCs,
    respectively.
    Both quantities are given as a function of $\Delta Q$ per adatom
    (lower axis) and $\Delta Q$ per surface area (upper axis).
    The solid/dashed lines and light/dark markers correspond to one-sided/two-sided adsorption of the two adatoms.
    Some $\Delta E$ data points are out of the scale,
    see Sec.~1.3 of the supplemental material\cite{Supplemental} for details.
    The schematic insets label the intermediate and resonant regions.
  }
  \label{fig:Emag1}
\end{figure*}

This subsection focuses on adatom arrays $\alpha$, $\beta$ and $\gamma$
(see Sec.~\ref{sec:methods:confs} on their definitions).
These structures have two adatoms in the unit cell and they are suitable for description of remote interactions.
Both one-sided adsorption and two-sided adsorption are considered.
The long range interactions between the adatoms depend on their respective lattice positions
and affect the magnetizations and energies of the FM and AFM solutions.
Figures~\ref{fig:Emag1}\,(a) and \ref{fig:Emag1}\,(b) illustrate the average
magnetization per adatom of the FM solutions
($\overline{M}=M(\text{FM})/2$)
for $6\times3$ and $8\times4$ graphene SCs,
respectively.
These quantities are functions of the average added charge per adatom ($\overline{\Delta Q} = \Delta Q/2$).
As an example,
the $\overline{M}$ curves of $\alpha(8\times4)$ are identical (but with fewer data points)
to the $M$ curves in Fig.~\ref{fig:4x4_mag_geom}
because the FM $\alpha(8\times4)$ solution is equal to doubling the solution of the $4\times4$ SC with one adatom.
The remaining $\overline{M}$ results are in overall similar.
At $\Vg=0$ the SCAN magnetizations are smaller than for PBE,
in fact for $\alpha(6\times3)$ it is even zero.
The slopes of the $\overline{M}$ curves depend on SC size as discussed in Sec.~\ref{sec:singleadatom},
but they do not depend significantly on the array type.
However,
there is a dependency on the offset magnetization so that at $\Vg=0$ the PBE magnetizations vary from
$\overline{M}(\alpha)=0.17\,\muB$ to
$\overline{M}(\gamma)=0.44\,\muB$
for $6\times3$ SCs
and from
$\overline{M}(\beta) =0.35\,\muB$ to
$\overline{M}(\gamma)=0.43\,\muB$ for $8\times4$ SCs.
The corresponding SCAN magnetizations vary from
$\overline{M}(\alpha)=0.00\,\muB$ to
$\overline{M}(\gamma)=0.40\,\muB$
for $6\times3$ SCs.
This variation of magnetizations can be connected to steep magnetization slopes and large DOS associated to the $\psip$ states.
In these conditions,
tiny energy shift produced by interactions of adatoms change $\psip$ occupations significantly.
For this reason,
the magnetization values are also sensitive to the computational details,
particularly on the choice of XC functional and density of the $k$-point mesh.
One can now understand why calculated magnetic moments vary significantly in the literature.\cite{Feng2017_review_2Dspintronics}
However,
with positive gate voltage the differences for the magnetization values become smaller.
Moreover,
$\overline{M}(\beta)$ and $\overline{M}(\gamma)$ robustly reach the saturation magnetism
at around $\overline{\Delta Q} = 1\,e$.

We turn now to discuss an important result of this study concerning
the energy differences $\Delta E = E(\text{FM}) - E(\text{AFM})$.
This quantity is plotted as a function of $\overline{\Delta Q}$
in Figs.~\ref{fig:Emag1}\,(c) and \ref{fig:Emag1}\,(d) for $6\times3$ and $8\times4$ SCs,
respectively.
The differences in $\Delta E$ between the respective one-sided and two-sided configurations are significant,
but differences in the magnetization are not.
We also observe very different energies in the different $\oDQ$ regions,
which have been discussed in Sec.~\ref{sec:methods:hamiltonian}.
In the intermediate region (with only partially filled $\psip$ impurity states),
PBE and SCAN behave differently,
but in both cases a clear ferromagnetic valley is formed in the middle of the  region for each array.
Representative $\Delta E$ maximum values are about $-30$\,meV and $-5$\,meV for $6\times3$ and $8\times4$ SCs,
respectively.
In the case of structures with two-sided adsorption
the energy difference magnitudes are larger for $\alpha$ but smaller for $\gamma$.
In most cases the total energy was found to be lower for the two-sided adsorption configuration
than for the respective one-sided one
(See Sec.~2 of the supplemental material\cite{Supplemental}).
For $\beta$ arrays,
the ferromagnetism for two-sided adsorption is clearly weaker for the $8\times4$ SCs
but for the $6\times3$ SCs there is no clear difference in the maximum $\Delta E$ magnitude.
In general,
the $\gamma$ arrays yield the most robust ferromagnetism,
especially for the $8\times4$ SCs.
Also the total energies are lower for the $\gamma$ arrays
(See Sec.~2 of the supplemental material\cite{Supplemental}).
The most stable FM solution depends on the occupation of the $\psip$ state.
Regardless of the SC size or the XC functional,
this solution corresponds to
$\overline{M}=(0.40\pm0.05)\,\muB$,
$(0.35\pm0.10)\,\muB$ and
$(0.55\pm0.05)\,\muB$
for $\alpha$, $\beta$ and $\gamma$ arrays, respectively.
These results can be understood within the Stoner model.
In the AFM solution there is no hybridization between $\psipup^1$ and $\psipdown^2$ impurity states,
which are associated to {\Cadyksi} and {\Cadkaksi} adatoms,
respectively.
On the contrary,
in the FM solution $\psipup^1$ and $\psipup^2$ hybridize,
which reduces the magnetization of the FM solution.\footnote{We have verified from our data that
the magnetic moments projected to {\Cadyksi} and {\Cadkaksi} 
are larger in the AFM solution than in the FM solution.} 
The hybridization leads to a trade for potential energy by the expense of kinetic energy. 
The Stoner's criterion reveals whether or not the trade was beneficial.
Because of the peak in the DOS,
the FM solution is stabilized.
This phenomenon explains also why the maximum of the FM interaction is given at $\overline{M}\approx0.5\,\muB$,
since the DOS reaches its maximum around half-filling.
Incidentally, 
the Stoner model becomes irrelevant in the resonant region,
where the impurity band is completely filled.
One can also visualize the present phenomenology as follows.
The hybridization between $\psipup^1$ and $\psipup^2$ creates bonding and antibonding states.
The filling of the bonding state ($\overline{M}\rightarrow0.5\,\muB$) stabilizes the FM solution,
while the filling of the antibonding state ($\overline{M}\rightarrow1\,\muB$)
weakens it.
When SCAN and PBE $\Delta E$ are compared,
the results in the most strongly FM region look similar,
suggesting that PBE captures the Stoner phenomenology rather well.
However,
the SCAN FM stability is systematically slightly weaker.
As a matter of fact,
SCAN promotes orbital localization,
since SCAN is more sensitive to chemical bonds than PBE by taking into account the kinetic energy density of the electrons.

For low magnetizations,
SCAN yields strong AFM excursions especially for $\gamma(6\times3)$ and $\beta(6\times3)$
but also for $\alpha(6\times3)$,
meanwhile this phenomenon is completely absent in the PBE results.
These AFM anomalies can be explained within conventional RKKY.
At low magnetizations,
the $\psip$ orbitals become more localized for SCAN,
but PBE cannot capture this localization.
As discussed in Sec.~\ref{sec:methods:hamiltonian},
the Stoner model can break down in this dilute limit.
In these conditions,
the indirect exchange mediated by the conduction graphene electrons dominates.
For low magnetizations,
the integration of Eq.~\eqref{eqExGen} has not yet reached the poles,
thus $\EexRKKY\approx\Eexconvbri$.
Now,
Eq.~\eqref{eq:Enonres} explains that $\Eexconv$ is slightly AFM for $\alpha(6\times3)$ and $\alpha(8\times4)$,
strongly AFM for $\beta(6\times3)$ and even more strongly AFM for $\gamma(6\times3)$.
This scenario qualitatively agrees with the SCAN results since strong AFM excursions appear
for $\gamma(6\times3)$ and $\beta(6\times3)$ and a weaker one for $\alpha(6\times3)$.
There is a small inconsistency with the conventional RKKY picture in the $\alpha(8\times4)$ curve,
where the expected small AFM peak is absent.
However,
as discussed in Sec.~1.3 of the supplemental material,\cite{Supplemental}
in this conventional RKKY region there is substantial numerical instability in the AFM solution,
manifesting the delicateness of the underlying physics.
Therefore,
$\Delta E$ is sensitive to computational details
and the SCAN results for low $\oM$ may contain error in the meV range.
Moreover,
it should be noted that the Eq.~\eqref{eq:Enonres} is only approximative for the bridge site
and that $\alpha(6\times3)$ and $\alpha(8\times4)$ arrays contain competing AFM and FM interactions.
Therefore,
the conventional RKKY ferromagnetism for $\alpha(8\times4)$ is not conclusive,
but the AFM peaks for $\gamma(6\times3)$ and $\beta(6\times3)$ are robust and
thus confirm the conventional RKKY model.

When the $\psip$ orbitals become fully occupied,
$\Delta E$ reaches AFM plateaus. 
In this region the impurities do not anymore interact via the Stoner's mechanism 
and the situation becomes similar to the Heitler-London limit for the hydrogen molecule,
where electrons on different sites have opposite spins.
However, 
the observed antiferromagnetism could be also explained by the RKKY formalism.
As discussed in Sec.~\ref{sec:methods:hamiltonian},
in this $\oDQ$ region RKKY is described within the resonant RKKY model,
which predicts AFM interaction energies because the impurity states lie below $\ED$.

\subsection{Close interaction adatom pair configurations}\label{sec:results:close}

\begin{figure*}
  \includegraphics[width=\linewidth]{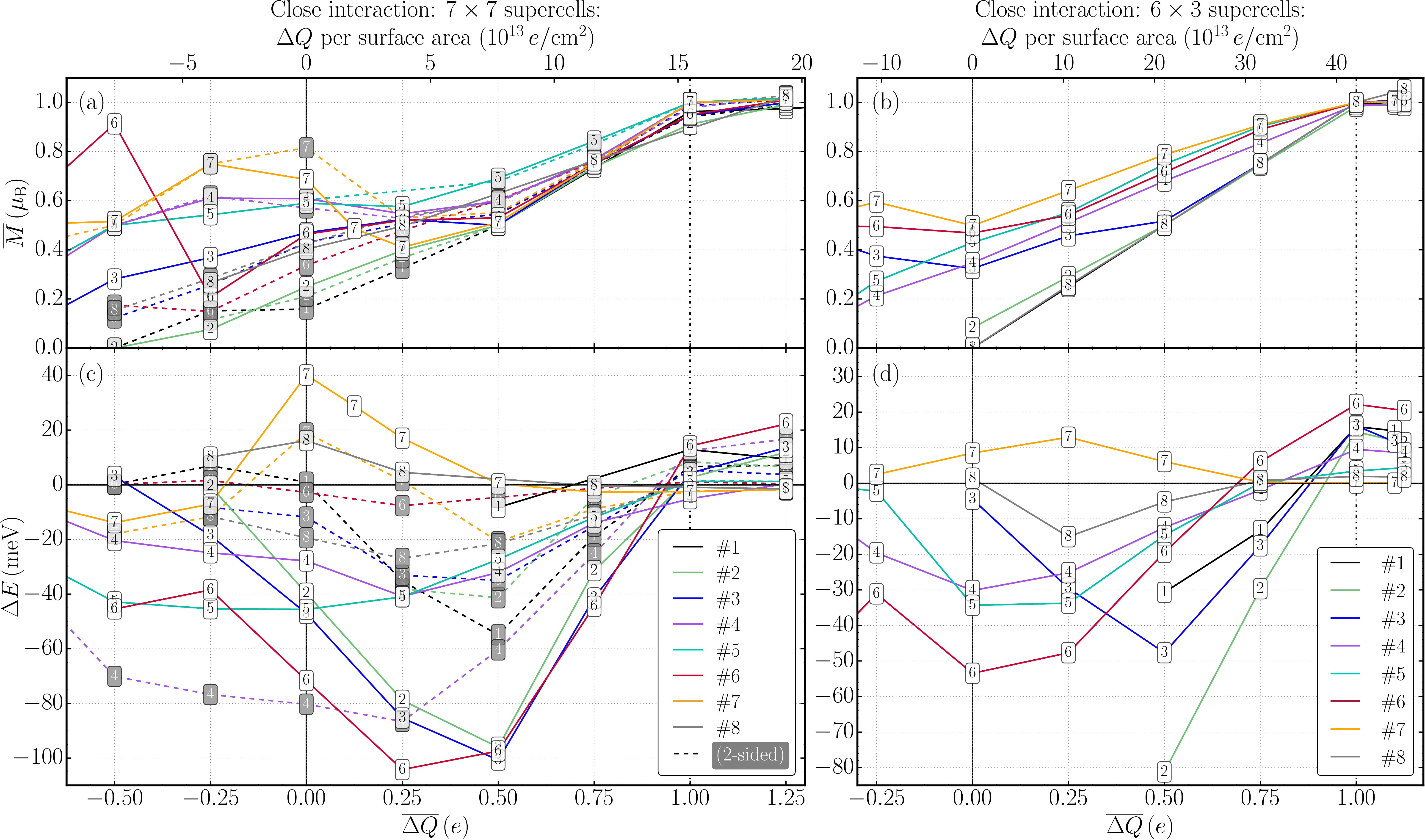}
  \caption{
    (Color online). 
    Close interaction pairs:
    panels (a) and (b) represent $\oM$ for $6\times3$ and $8\times4$ SCs,
    respectively.
    Panels (c) and (d) contain the $\Delta E = E(\text{FM}) - E(\text{AFM})$ values for $6\times3$ and $8\times4$ SCs,
    respectively.
    Both quantities are given as a function of $\oDQ$ (lower axis)
    and $\Delta Q$ per surface area (upper axis).
  }
  \label{fig:Emag2}
\end{figure*}

We now consider that the adatoms are in the neighborhood of each other
by using close interaction pair Confs.~\#1--\#8 in $7\times7$ and $6\times3$ SCs.
A striking difference with respect to the remote interaction cases are non-linear growths of the magnetizations curves
displayed in Figs.~\ref{fig:Emag2}\,(a) and (b).
The $\Delta E$ values shown in Figs.~\ref{fig:Emag2}\,(c) and (d)
are also substantially larger than the corresponding remote interaction energies.

Let us first focus on the $7\times7$ SC structures,
which represent well-isolated adatom pairs.
In the region $\overline{\Delta Q} \lesssim 0.5\,e$,
the magnetizations stay roughly constant or even decrease when more electrons are added.
This behavior is produced by minority spin $\psippdown$ orbitals.
Normally,
these states lie deep in energy (as shown in Fig.~\ref{fig:dos}),
but strong interactions between adatoms and large structural distortions raise these states to $\EF$
(see the DOS plot in Sec.~3.2 of the supplemental material\cite{Supplemental}).
Upon increasing $\overline{\Delta Q}$ both $\psipup$ and $\psippdown$ are being occupied.
These fillings yield opposing contributions to the magnetization.
Thus increasing $\overline{\Delta Q}$ often results in zero or even negative magnetization slope.
Moreover,
the $\psipp$ states may have different occupations at {\Cadyksi} and {\Cadkaksi} at low $\Delta Q$ values,
leading to adatom magnetic moments with different magnitudes.
Thus in many cases the AFM solution becomes
a ferrimagnetic solution with nonzero total magnetization
(see Sec.~3.2 of the supplemental material\cite{Supplemental} for details about these cases).
Furthermore,
the $\psipp$ states bring complication in $\Delta E$ because the number of interacting states becomes higher
and PDOS($\EF$) can have both significant spin up and spin down components.
Besides,
the graphene Dirac cone is severely deformed.
Nevertheless,
the values for $\Delta E$ reveal a preference for strong ferromagnetism with values as high as $-104$\,meV (Conf.~\#6)
but Confs.~\#7 and \#8 display antiferromagnetism (or ferrimagnetism) with maximum magnitude of 40\,meV.
These results suggest high Curie temperatures in many of the studied configurations.

We now discuss higher gate voltages.
Beyond the limit $\overline{\Delta Q}\gtrsim0.5\,e$,
the $\psipp$ states become fully occupied.
In this regime,
the magnetization and $\Delta E$ recover a behavior similar to the remote interaction cases discussed in the previous section.
In the intermediate region ($0.5\,e \lesssim \overline{\Delta Q} \lesssim 1\,e$),
the magnetizations are linear and their values for the respective one-sided and two-sided adsorption models are almost equal.
Moreover,
each one stabilizes the FM solution (though some cases only weakly).
Above $\oDQ=1\,e$ the close interaction configurations behave similarly with the adatom array ones. 
The magnetizations saturate and the antiferromagnetism is stabilized.  
In this case the RKKY theory is inapplicable 
because the Dirac cone is deformed and the rigid band approximation has become invalid.

The $6\times3$ SC structures have significant next-nearest neighbor interactions between
{\Cadyksi} and the periodic images of {\Cadkaksi},
and vice versa.
All these interaction distances are in the range {5\,\AA--\,10\,\AA},
which corresponds to the remote interaction case with FM character in the region $\oDQ<1\,e$.
Thus,
in comparison to the $7\times7$ SC results,
one expects to observe stronger ferromagnetism in Confs.~\#1--\#6
and weaker antiferromagnetism or even ferromagnetism in the Confs.~\#7 and \#8.
Actually,
Conf.~\#8 becomes FM and \#7 less AFM.
On the contrary,
the ferromagnetism in Confs.~\#1--\#6 decreases as typical $\Delta E$ values
drop to around $-30$\,meV and are about $-55$\,meV at maximum.
The reason is that the Stoner criterion is weakened because
the tighter packing of the adatoms spreads the $\psip$ energy band.
The typical widths of the $\psip$ states in PDOS are $\sim0.4$\,eV but for the corresponding
$7\times7$ SC configurations they are $\sim0.1$\,eV.
In our tests with even tighter packing the ferromagnetism is further reduced.
This behavior is consistent with Ref.~\onlinecite{Edwards2006_sp_stoner_ferromag},
which argues that in band ferromagnetism with $sp$ character the magnetization must be inhomogeneous with
only a fraction of the sample ferromagnetically ordered.
In the case of fully filled $\psip$ states the results are nearly equal
with respect to the $7\times7$ SC cases.

The total energies of each structure have been listed in
Sec.~2 of the supplemental material.\cite{Supplemental}
Regardless of the SC size,
the Conf.~\#4 was found to be lowest in energy and Conf.~\#6 the highest,
with an energy difference of 337\,meV between these cases for the one-sided adsorption at $7\times7$ SC.
Adsorbing the adatoms to the different sides of the graphene sheet rather than on one side
was found to be energetically favorable in most cases,
as in the case of adatom arrays.
The $7\times7$ SC Conf.~\#4 was found to be $5.42$\,eV higher in energy with respect to the dimerized solution,
suggesting the existence of strong attractive potential between the adatoms.
As a matter of fact,
the $7\times7$ SC Conf.~\#4 was found to have 314\,meV lower energy than two individual adatoms on $7\times7$ SCs.
This suggests that a low temperature might be needed to stabilize the studied adatom pairs.

In summary,
our close interaction results show robustness of the Stoner paradigm for the C adatom system on graphene.
Even when the $\psipp$ states are present at $\EF$ (below $\overline{\Delta Q}\approx0.5\,e$),
the general trend is that the interaction energies are FM in the intermediate region,
which implies that the ferromagnetism could persist in more complicated
and experimentally realizable carbon systems.\cite{Kim2014}
The gate voltage can be used to further optimize FM interaction strengths.
Our study also reveals that in this regard $6\times3$ is roughly the optimal SC size for two adatoms,
which corresponds to one adatom per 18 graphene atoms.
However,
our results indicate that within this concentration,
it is beneficial if adatoms are scattered inhomogeneously,
since stronger ferromagnetism is observed for close interaction adatom pairs than for adatom arrays
within the $6\times3$ SCs.

\subsection{The spin polarization maps}

\begin{figure}
\centering
\includegraphics[width=.9\linewidth]{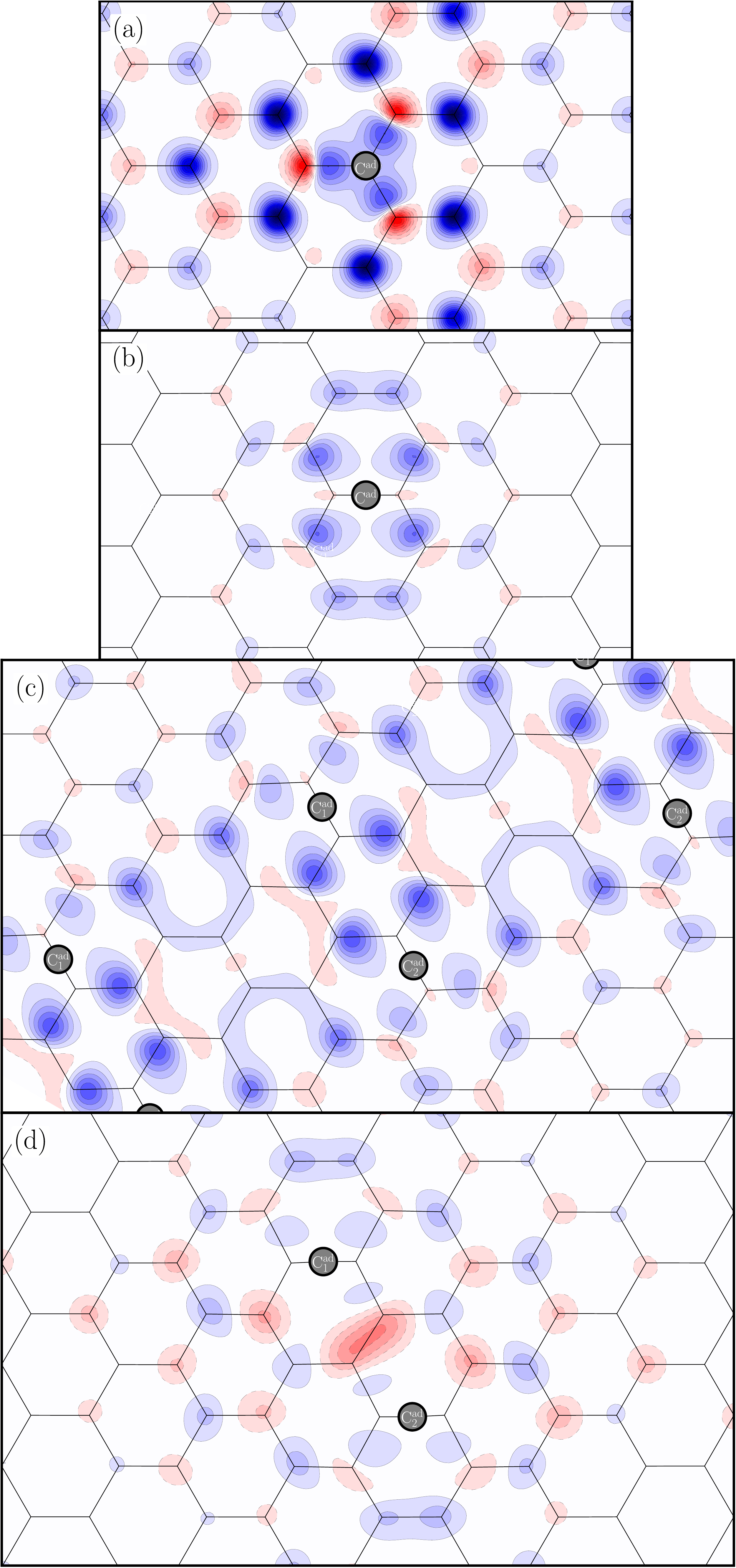}
\caption{\label{fig:contours}
  (Color online).
  Spin polarization maps with $\Vg = 0$ for
  (a): top site ($7\times7$ SC),
  (b): bridge site ($7\times7$ SC),
  (c): FM solution of Conf.\,\#2 ($6\times3$ SC),
  (d): FM solution of Conf.\,\#4 ($7\times7$ SC).
  Blue with solid contour lines and red with dashed contour lines denote
  positive and negative magnetization densities,
  respectively.
  The isovalues are every $0.002\,e/\AAmath^3$ except for (a),
  for which this value is $0.004\,e/\AAmath^3$.
  The data is taken at a slice about $0.4\,\AAmath$ below the graphene sheet
  (at the opposite side of the adatoms).
}
\end{figure}

Experimentally,
the adatom-induced spin polarization of the graphene sheet can be probed by using the scanning tunneling microscopy (STM),
as shown by Gonz\'alez-Herrero \emph{et al.}\ for hydrogen adatoms.\cite{GonzalezHerrero2016}
The DFT calculations also yield these spin patterns and they can be visualized as spatial spin polarization maps.
Figure~\ref{fig:contours} illustrates these maps for selected C adatom configurations at $\Vg=0$.

Such map for a spin up C adatom fixed on the top site at $A$ sublattice ($M=2.00\,\muB$)
is shown in Fig.~\ref{fig:contours}\,(a).
The magnetization pattern has triangular symmetry and the magnetizations at
$A$/$B$ sublattice sites are always positive/negative,
respectively.
The polarization is caused by the impurity state which extends to the graphene lattice
and by the polarized graphene electrons carrying the indirect interactions.
The tail of the impurity state is composed of $p_z$ orbitals
belonging to the $A$ sublattice sites.
The $B$ sublattice $p_z$ orbitals are not involved
due to the bipartite nature of the graphene lattice.
The corresponding patterns are similar for many magnetic defects coupling only to one sublattice,
e.g. substitutional transition metal atoms.\cite{Crook2015,Thakur2017}
The $B$ sublattice sites are negatively polarized because the spin up impurity state at
the $A$ sublattice sites attracts spin up electrons from the $B$ sublattice sites,
thus magnetizing them negatively.
In addition another impurity state placed on the $B$ sublattice interacts with the spin down graphene states.
The resulting indirect interaction with the original $A$ sublattice impurity is AFM
in agreement with the Lieb's theorem.
However,
in the case of the hydrogen adatom,
the magnetization pattern is inverted
(the $A$/$B$ sublattice sites have negative/positive magnetizations, respectively)\cite{Casolo2009_H-adatoms,Sofo2012,GonzalezHerrero2016}
as explained by Casolo \emph{et. al.}\cite{Casolo2009_H-adatoms}

The richness of the DFT simulations is also reflected in the case of bridge configurations.
Figure~\ref{fig:contours}\,(b) presents the magnetization density associated with a single
adatom on a bridge site ($M=0.43\,\muB$).
Our simulations indicate that the corresponding magnetic patterns are general
for any magnetic impurities coupled equally to the both graphene sublattices.
The amplitudes are mostly positive,
in particular,
one can notice four blobs between the basal atoms
and their nearest neighbor graphene atoms and six smaller magnetization regions,
two of which extend to two graphene sites.
These regions are also visible in the spin polarization isosurface plot in Fig.~\ref{fig:vmd}.
Further away positive and negative magnetizations alternate but positive amplitudes dominate
similarly with the top-site situation.
The magnitudes of the polarizations are strongly reduced when compared to the top site case
for two possible reasons.\footnote{Notice that the contour isovalues
are twice as large for the top site plot in Fig.~\ref{fig:contours}\,(a)
than in the bridge site plots.}
Firstly,
the $\psip$ state now spreads both to the $A$ and $B$ sublattices.
Secondly,
the magnetization of the C adatom is much higher for the top site configuration.
In fact,
by increasing gate voltage,
we observe strengthening of the positive magnetizations and gradual disappearance of the negative magnetizations.
The resulting pattern is highly asymmetric.
Consequently,
the direct coupling acquires a strong angular dependency to the vector connecting the adatoms.
Moreover,
the respective angle of the basal sites of {\Cadyksi} and basal sites of {\Cadkaksi}
(which can be {$\theta_\text{bas}=0$\textdegree}, {60\textdegree} or {120\textdegree})
becomes relevant.
The present results show an overall trend for stronger ferromagnetism for the {$\theta_\text{bas}=0$\textdegree} angle,
as in the case of $\gamma$ arrays and Confs.~\#2, \#4, \#5, and \#6.
One might
achieve adsorption with {$\theta_\text{bas}=0$\textdegree}
by applying uniaxial strain,
leading to longer bonds in one direction.
The strain might also stabilize the adatom network.
These properties of the strained system are of interest for future study.

Figure~\ref{fig:contours}\,(c) visualizes the magnetization patterns
for non-isolated Conf.~\#4 adatom pair ($\oM=0.70\,\muB$),
which corresponds to a FM interaction with $\Delta E = -30\,$meV.
In this case,
the positive magnetizations induced to graphene by the two adatoms mutually strengthen,
thus a mainly positive magnetization develops on the whole unit cell.
This effect can be also pictured as a stronger delocalization of the $\psip$ states.
In this $\Delta Q = 0$ plot,
negative polarization areas are still visible,
however,
they gradually disappear when more charge is added.

Figure~\ref{fig:contours}\,(d) contains magnetization patterns for
isolated Conf.~\#2 adatom pair ($\oM=0.50\,\muB$),
which corresponds to a FM interaction with $\Delta E = -40\,$meV.
An interesting feature is the negative excursion of the magnetization in between the adatoms,
which is much stronger than any positive magnetization in the contour map,
even though the positive magnetizations typically dominate in the FM solutions,
as shown in Fig.~\ref{fig:contours}\,(c).

Although magnetization patterns of hydrogen have been recently observed
by Gonz\'alez-Herrero \emph{et al.},\cite{GonzalezHerrero2016}
at the moment no experiments have yet measured the polarization maps related to bridge configurations.
Since our DFT calculations predict that these maps are completely different,
STM experiments probing these cases would be useful to test our present models.

\section{Conclusions}\label{sec:conclusions}

By using DFT first principle simulations and various models of magnetism,
we have proposed how to use gate voltage to control magnetism of carbon graphene adatoms.
We have found that the voltage bias can be deployed to fine tune the magnetic moment of the adatoms
from zero to $1\,\muB$ by emptying or filling the localized magnetic state on the adatom.
Moreover,
the gate voltage influences the strength and sign
(FM or AFM) of the magnetic interactions between the adatoms.
We find AFM behavior at both low and high adatom magnetic moment regions and ferromagnetism
at intermediate adatom magnetic moments.
At this intermediate region the Fermi level falls in the band of impurity states 
which have formed by direct hybridizations between the impurity states despite the relatively
large adatom separations (up to 1\,nm).
As a result,
the adatom spins become ferromagnetically ordered since the Stoner's criterion is fulfilled. 
The strong low-magnetization antiferromagnetism can be explained within the RKKY mechanism.
This antiferromagnetism was only found simulations within the more accurate SCAN meta-GGA framework,
which is capable of confining the impurity states better,
eliminating the direct exchange for low magnetic moments. 
The antiferromagnetism at high magnetic moments could result from an
interplay between direct and RKKY interactions.
In this case,
the RKKY interaction is described within the novel resonant RKKY model,
which predicts AFM interaction energies because the impurity states lie below $\ED$.
The existence of different regimes which can be switched by Fermi level variation
highlights that the system is more complex than previously believed.
Previously only either (generalized) indirect RKKY exchange or direct exchange
has been considered in the case of one system.

The DFT calculations reveal that both GGA and SCAN predict strongly FM configurations
in the intermediate magnetic moment region,
nevertheless there is a small reduction when SCAN is applied.
Calculated spin polarization maps show highly non-trivial magnetic distributions in space,
which could be probed with magnetic scanning microscopes.
The present results can be generalized
to other systems with magnetic defects which couple to the both $A$ and $B$ graphene sublattices
and can explain some experimental findings concerning magnetism in systems based on graphene,
useful for future spintronics applications.

\begin{acknowledgments}

We are grateful to Prof. A. Feiguin,
Prof. A. Bansil and C. Lane for useful discussions.
This work was supported by the Academy of Finland Grants No.~277829, 318500
and benefited from the generously allocated computer time at CSC -- Finnish IT Center for Science.
\textsc{pymatgen} package\cite{pymatgen} was employed in plotting and data analysis.

\end{acknowledgments}

\bibliography{bibliography}

\end{document}